# A new family of bioSFQ logic/memory cells

Vasili K. Semenov[1], Evan B. Golden[2], and Sergey K. Tolpygo[2]

[1] Department of Physics and Astronomy, Stony Brook University, Stony Brook NY 11794
[2] Lincoln Laboratory, Massachusetts Institute of Technology, Lexington, MA 02421

*Abstract*—Superconductor electronics (SCE) is competing to become a platform for efficient implementations of neuromorphic computing and deep learning algorithms (DLAs) with projects mostly concentrating on searching for gates that would better mimic behavior of real neurons. In contrast, we believe that most of the required components have already been demonstrated during the long history of SCE, whereas the missing part is how to organize these components to efficiently implement DLAs. We propose a family of logic/memory cells in which stored multi-bit data are encoded by quasi-analog currents or magnetic flux in superconductor loops while transmitted data are encoded as the rate of SFQ pulses. We designed, fabricated, and tested some of the basic cells to demonstrate a proof of concept, e.g., a unipolar and bipolar multipliers based on Josephson junction comparators. We coined the term bioSFQ to clearly connote close but distinguishable relations between the conventional SFQ electronics and its new neuromorphic paradigm.

## I. Introduction

SUPERCONDUCTOR electronics (SCE) has demonstrated the lowest energy dissipation per operation [1] and the highest clock rates (~770 GHz) [2] among existing electronics. It is more mature than other beyond-CMOS technologies for classical and quantum computing. A number of hardware technologies are competing to become a platform for energy-efficient artificial neural networks (ANNs), including conventional CMOS (see, e.g. [3]), more exotic memristor-based [4], and perhaps even more exotic superconductor-based technologies [5], [6].

Josephson junctions (JJ) can act as natural spiking neuron-like devices for neuromorphic computing [7]. Most of the existing projects on implementation of superconductor electronics for neuromorphic computing concentrate on searching for gates that would better mimic behavior of real neurons, e.g., using adiabatic quantum flux parametrons [8], flux qubits, phase slip junctions [9], magnetic JJs [10], single photon detectors [11], [12], etc. [13]. Many JJ-based suggestions are purely theoretical and only a few operational JJ-based ANNs have been reported [13]-[16].

We believe that most of the required components have already been invented (in particular, by our team members) for superconductor digital electronics, whereas the essential missing part is how to creatively utilize and organize (design and fabricate) these components in large-scale, fast, and energy-efficient artificial neural networks. In this work, we propose a family of logic/memory cells in which stored multi-bit data are encoded by quasi-analog currents or magnetic flux in superconductor loops while transmitted data are encoded as the rate of SFQ pulses. We coined the name "bioSFQ" to clearly connote a close but distinguishable relation between the conventional SFQ electronics and its new neuromorphic paradigm. In the proposed devices and circuits, information propagates with the speed of light and its quasi-stochastic processing naturally emulates electrical processes in the brain.

Unique features of SCE continue to keep it attractive for large-scale electronic systems. However, in all prior attempts to implement SCE for general-purpose or high-performance computing it mimicked well-developed CMOS approaches and, as a result, its powerful potential has not been unleashed. Neuromorphic computing could become the greatest exception.

## II. Basic Components

### A. Storage of Analog Bipolar Data

Closed superconductor loops are traditional devices for storage of quantized magnetic flux. Josephson junction circuits allow the control the flux with single flux quantum accuracy as discussed, for example, in [17], Fig. 4. We implement such storage component as relatively long, thin film strips placed over one or between two ground planes; see Fig. 1c. The edges of the strip are terminated by flux pumps or, if necessary, by another Josephson junction-based circuitry.

### B. Reading of Analog Bipolar Data

The value of current can be measured by a Josephson current comparator; see, for example, [18]. However, in the case of a superconducting current in a loop, it is more natural to use a non-destructive comparator or C-SQUID [19] that does not affect the measured current, as shown in Fig. 1a.

This research is based upon work supported by the Under Secretary of Defense for Research and Engineering under Air Force Contract No. FA8702-15-D-0001. *(Corresponding author: Vasili Semenov.)*

V. K. Semenov is with the Department of Physics and Astronomy, Stony Brook University, Stony Brook, NY 11794-3800, USA (e-mail: Vasili.Semenov@StonyBrook.edu).

E. B. Golden and S. K. Tolpygo are with the Lincoln Laboratory, Massachusetts Institute of Technology, Lexington, MA 02421, USA (e-mails: Evan.Golden@ll.mit.edu, Sergey.Tolpygo@ll.mit.edu). Color versions of one or more of the figures in this paper are available online at http://ieeexplore.ieee.org.

Digital Object Identifier will be inserted here upon acceptance.

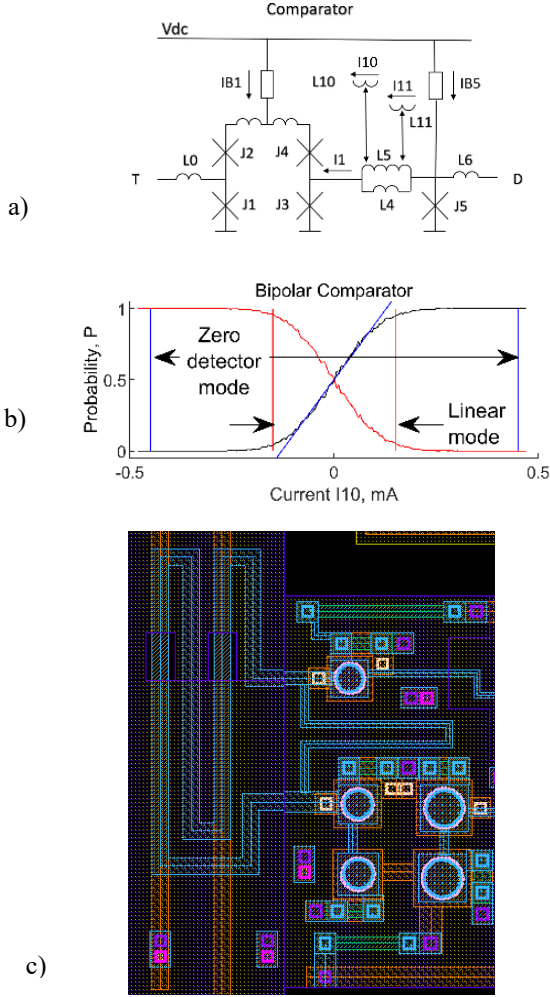

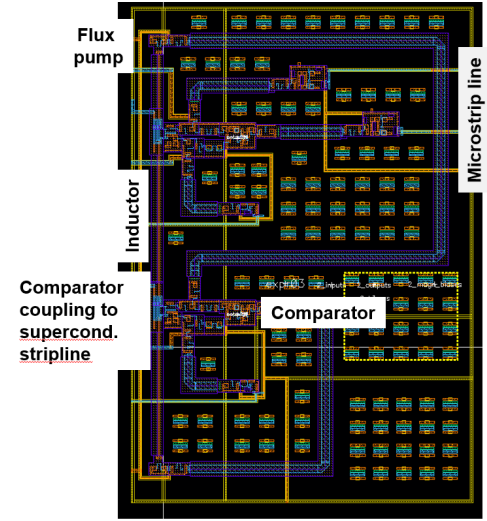

Fig. 2. A sparsely designed fragment (220 μm × 320 μm) of neuromorphic unit cell showing comparators coupled to superconducting storage loops, flux pump, and superconducting transmission lines for transferring SFQ pulses between distant parts of the circuit.

Fig. 1. Analog circuitry for bipolar current measurements: a) circuit diagram of a C-SQUID; b) its transfer function and operation; c) a fragment of the layout with circular Josephson junctions and inductors L10 and L11 (red vertical strips) that can be parts of superconducting multi-flux storage loops described in II.A.

The comparator converts the applied current, $I$ into the rate of SFQ pulses. The probability of an SFQ pulse passage through the comparator is given by the error function

$$p(X) = \tfrac{1}{2}\big(1 + \mathrm{erf}(\pi^{1/2}X)\big), \quad (1)$$

$$X \equiv (I - I_{th})/\Delta I, \quad (2)$$

where $I_{th}$ is the (adjustable) threshold current of the comparator. If current $I$ is the comparator input current $I1$ shown in Fig.1a, then the width of the comparator gray zone $\Delta I$ is defined by the thermal and quantum noise in the comparator junctions [18] – [22]. If $I$ is the current in the primary of the C-SQUID transformer L10, then the effective gray zone $\Delta I$ is a custom-adjustable parameter defined by the transformer L5-L10 to the desired effective gray zone width, e.g., −0.15 mA to 0.15 mA in Fig. 1b.

The black trace in the Fig. 1b shows the measured dependence of the probability of the SFQ pulse transmission on current flowing via inductor L10. The desired widening of the grey zone is achieved by the additional attenuation of the current induced in L5 by shunting inductor L4. As a result, the inductive divider composed of L5 and L4 reduces the feeding current $I1$ injected between comparator junctions J3 and J4. The red trace in Fig. 1b shows the complementary probability of no SFQ response on the clock pulse. The complementary output in the circuit is calculated by passing the direct output through an RSFQ NOT cell, not shown in Fig. 1.

The comparator transmission function (1) is nonlinear and saturates at $|X| > 1$. This behavior closely resembles the sigmoid response function frequently used to describe behavior of neurons. In general, saturation of the response function with stimulus strength is typical for all biological systems. Hence, this feature of the comparator response (1) is highly beneficial for neuromorphic applications.

Near the threshold, $|X| < \Delta I$, expansion of (1) is very close to the linear response

$$p(X) = \tfrac{1}{2} + X, \quad (3)$$

shown as the black straight line in Fig. 1b. The linear mode of operation allows us to perform many useful operations in a very simple manner, see III below.

*C. Other Components*

Layout of a simple bioSFQ circuit comprising a few basic components is shown in Fig. 2. In fact, all basic components of bioSFQ were reviewed in [23] and many later publications as components of SFQ digital circuits. After 30 years in existence, some of them were "reinvented" as neural network cells; for example, the splitter cell was reintroduced in [24].

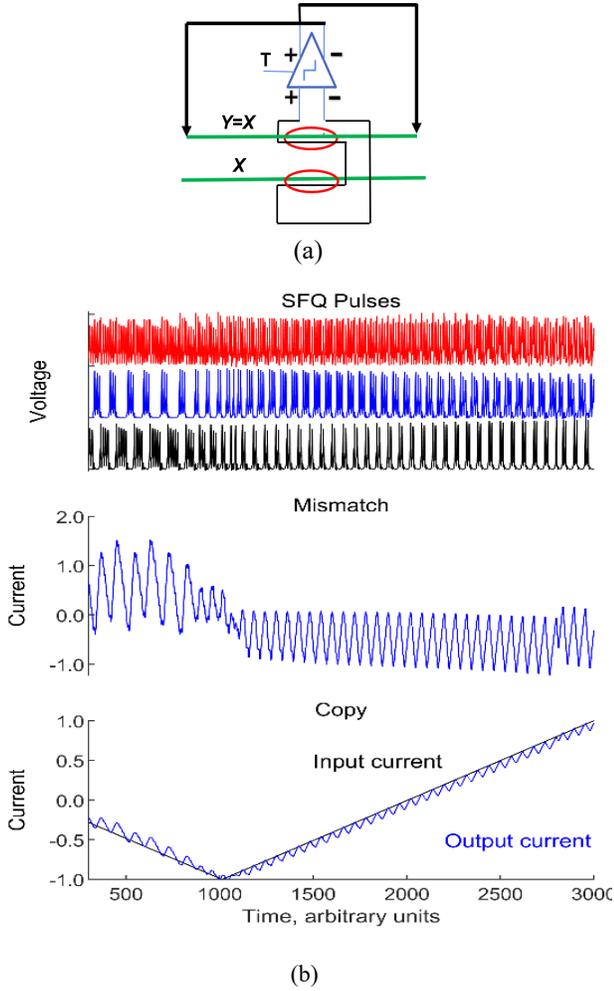

Fig. 3. Block diagram (a) and numerical simulation (b) of bipolar copying of current. Three upper traces are clock pulses (red), negative (blue) and positive (black) outputs of the comparator.

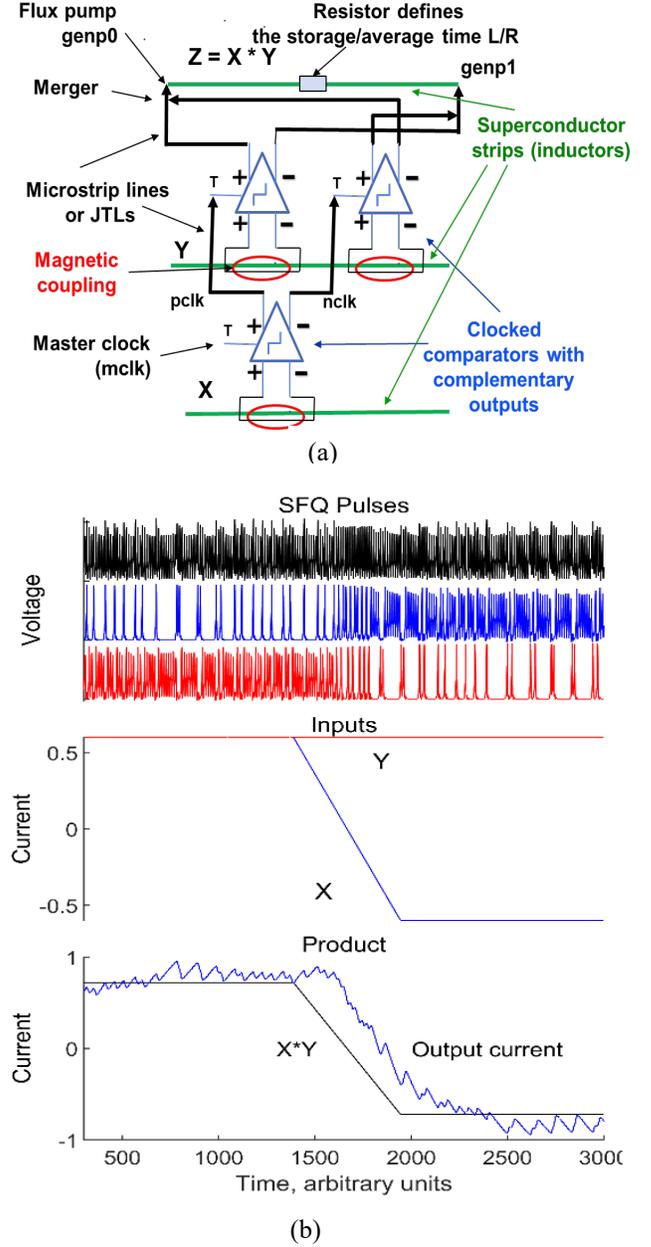

Fig. 4. Block diagram (a) and simulated operation (b) of bipolar multiplier. Upper traces show master clock (black color) pulses applied to the lower comparator, negative (blue) and positive (red) SFQ pulses applied to line $Z$.

## III. BASIC FUNCTIONS

### A. Copying of Bipolar Currents

Two inputs in Fig. 1a (inductor L10 and L11) are required to use the comparator for measuring a linear combination of currents $I10$ and $I11$ flowing in the corresponding inductors. In particular, if mutual inductances of L5 and L10, and of L5 and L11 are equal and have opposite signs $M_{L5,L10} = -M_{L5,L11}$, the comparator measures difference of the two input currents, as shown in Fig. 3. Trains of "positive" and "negative" SFQ pulses shown in the upper traces in Fig. 3b are continuously applied to opposite ends of inductor $Y$, as shown in Fig. 3a. As a result, the cell is trying to keep the difference between the signals (the middle trace in Fig. 3b) as low as possible. The lower traces show results of the simulation of the Copy operation: copying current $X$ (black trace) in one inductor to current $Y = X$ (blue trace) in another inductor.

### B. Bipolar Multiplication

A bipolar multiplier requires three comparators operating in the proportional mode; see Fig. 4. The first (lower) comparator in Fig. 4a converts operand $X$ to positive and negative trains of SFQ pulses which serve as clock pulses for two upper $Y$ comparators. As a result, the rates of output SFQ pulses from $Y$ comparators are proportional to a product of $X$ and $Y$. Two mergers join the corresponding trains of pulses and inject them to an inductor with large inductance. Earlier we used superconductor loops that store data for an infinitely long time, even after interruption of all SFQ pulses. Here we demonstrate a solution for a temporal storage of data. A relaxation of current to zero value is provided by a small resistor shown on top of Fig. 4a. The relaxation time is defined by $L/R$ constant. At no-zero $Z$, the rates of positive and negative pulses are unbalanced and the value of current is proportional to the difference of positive and negative rates of SFQ pulses applied to the opposite edges of line $Z$.

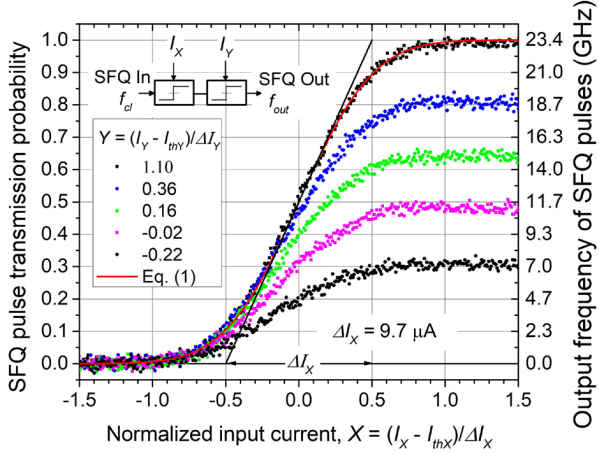

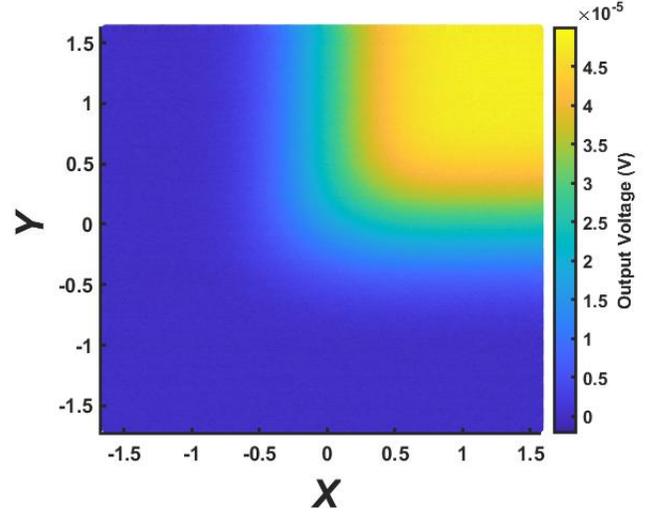

Fig. 5. Testing of the unipolar multiplier comprising two comparators connected in series. The time-averaged output voltage, $V_{out}$ of the device was measured and converted into the SFQ output frequency using the Josephson relation $f_{out} = V_{out}/\Phi_0$. The SFQ pulse transmission probability is calculated as $f_{out}/f_{cl}$. The output was measured at different input currents to the second comparator. Theoretical dependence (1) fit is shown by the red curve for the device at $Y = 1.1$ when the transmission probability of second comparator is 1 and Eq. (2) reduces to Eq. (1). The gray zone of the first comparator $\Delta I_1$ from the fit is 9.7 µA, which is in the excellent agreement with the theory in [22] and prior measurements [18], [20], [21]. Inset shows the block diagram of the multiplier.

Fig. 6. Color-coded map showing the output voltage (time-average of the output SFQ pulses) of the unipolar multiplier shown in Fig. 5 as a function of normalized applied currents to the first and the second comparators: $X = (I_x - I_{thX})/\Delta I_x$ and $Y = (I_y - I_{thY})/\Delta I_y$. Zero probability of transmission of the input SFQ pulses through the two comparators corresponds to blue color, and transmission with probability 1 corresponds to bright yellow. The input SFQ frequency $f_{cl}$ =23.4 GHz.

The middle traces in Fig. 4b show variations of input signals $X$ and $Y$ with time. $Y$ remains positive the entire time, while $X$ changes from a positive to a negative value The lower traces in Fig. 4b show the idealized product $X \cdot Y$ (black color) and the actual (simulated) multiplication results produced by the circuit (blue color).

### C. Demonstration of Unipolar Multiplication

Demonstration of the bipolar multiplier is in progress. At this point in time, we present the measured properties of a unipolar counterpart. It requires just two comparators connected in series as shown in Fig. 5. The multiplier utilizes the fact that the probability of independent events equals the product of individual probabilities. The first comparator ($X$) is fed by SFQ pulses with frequency $f_{cl}$ (from the clock source) and converts them into a stream of output SFQ pulses with frequency $f_x = p_X f_{cl}$. The SFAQ pulse transfer probability $p_X$ is given by (1), (2) and defined by the applied current $I_X$, the corresponding threshold current $I_{thX}$, and the width of the gray zone $\Delta I_X$. Similarly, the second comparator ($Y$) multiplies its input frequency $f_x$ by the its transfer probability $p_Y$, which is defined by the measured current ($I_Y$) and the corresponding parameters $I_{thY}$ and $\Delta I_Y$ of the second comparator.

The output frequency of the two-comparator multiplier is given by

$$f_{out} = p_x p_y f_{cl}. \quad (4)$$

The time-averaged output voltage of the device is given by the Josephson frequency to voltage relation $V_{out} = \Phi_0 f_{out}$. In the linear regime of both comparators, (4) reduces to

$$f_{out} = f_{cl}\left(\frac{1}{2} + X\right)\left(\frac{1}{2} + Y\right). \quad (5)$$

The measurement results of the described two-operand unipolar multiplier are shown in Figs. 5 and 6. The circuit was fabricated in the 8-Nb layer planarized process SFQ5ee using Nb/Al-AlO$_x$/Nb junctions with Josephson critical current density of 100 µA/µm$^2$ and the minimum linewidth of 0.25 µm [25]. We have found that both comparators are very well described by (1), and the output of the multiplier is described by (4) at up to 35 GHz sampling frequency.

The full testing results are demonstrated in the three-dimensional color map in Fig. 6 showing the color-coded normalized output frequency as a function of both input currents $I_X$ and $I_Y$. One can see that the device is fully functioning and operates as expected. Obviously, the demonstrated device can be extended to unipolar multiplication of multiple operands by adding more comparators in series.

## IV. CONCLUSION

We presented quasi-analog Josephson junction circuits which can be useful for the implementation of deep learning algorithms and neuromorphic computing. The circuits could be easily modified for large fan-ins and fan-outs because long storage inductors could be coupled with many comparators and many trains of SFQ pulses could be split and merged if necessary. Although presented and demonstrated circuits are composed of known components, their parameters are highly original. More importantly, the circuits serve as demonstrations of the potential of other prospective quasi-analog superconductor circuits.

Our near-term goal is in scaling up the suggested bioSFQ approach to large-scale circuits utilizing the availability of a unique fabrication technology at MIT LL, sufficient to manufacture circuits with up to ten million Josephson junctions per chip [25], [26] as well as a recently demonstrated AC/SFQ biasing technique of SFQ circuits [27].

However, our dream project is beyond implementation of the known deep learning algorithms which, as far as we know, are quite disconnected from the known successful implementations of deep learning in nature, *e.g.*, as humans. This is because human brains do not contain multi-bit adders and multipliers required for implementing machine learning algorithms. In this respect, our bioSFQ approach is not going to change this. We share the known opinion that the cortex effectiveness depends on the complexity of the circuit rather than on the details of neuron interconnections. It would be fantastic to create superconductor cortex-like circuitry so complex that it would start behaving as a real cortex.

ACKNOWLEDGMENT

We are grateful to Alex Wynn for his interest in and support of this work. The numerical simulations were performed using PSCAN2 software package developed by Pavel Shevchenko [28]. We thank to Coenrad Fourie for assistance with InductEx software [29] used for inductance extraction from circuit layouts. We are also grateful to Vlad Bolkhovsky and Ravi Rastogi for overseeing the wafer fabrication.

This material is based upon work supported by the Under Secretary of Defense for Research and Engineering under Air Force Contract No. FA8702-15-D-0001. Any opinions, findings, conclusions or recommendations expressed in this material are those of the author(s) and do not necessarily reflect the views of the Under Secretary of Defense for Research and Engineering. Delivered to the U.S. Government with Unlimited Rights, as defined in DFARS Part 252.227-7013 or 7014 (Feb 2014). Notwithstanding any copyright notice, U.S. Government rights in this work are defined by DFARS 252.227-7013 or DFARS 252.227-7014 as detailed above. Use of this work other than as specifically authorized by the U.S. Government may violate any copyrights that exist in this work.